\documentclass[12pt,preprint]{aastex}

\usepackage{epstopdf}

\newcommand{\mef}{M85\,OT2006-1}
\newcommand{\vete}{V838~Mon}

\def\farcs{\hbox{$.\!\!^{\prime\prime}$}}  				

\def\lsun{L$_{\odot}$}
\def\rsun{R$_{\odot}$}
\def\asec{\ifmmode ^{\prime\prime}\else$^{\prime\prime}$\fi}

\shorttitle{Spitzer Observations of M85\,OT2006-1}
\shortauthors{Rau et al.}

\begin{document}


\title{Spitzer Observations of the New Luminous Red Nova M85\,OT2006-1}


\author{A. Rau\altaffilmark{1}, S.R. Kulkarni\altaffilmark{1}, E.O. Ofek\altaffilmark{1}, L. Yan.\altaffilmark{2}}
\affil{$^1$ Caltech Optical Observatories, MS 105-24, California Institute of Technology, Pasadena, CA 91125, USA}
\affil{$^2$ Spitzer Science Center, California Institute of Technology, Pasadena, CA 91125, USA}

\email{arne@astro.caltech.edu}

\begin{abstract}

\mef\ is the latest and most brilliant addition to the small group of known Luminous Red Novae (LRNe). An identifying characteristic of the previously detected events (M31~RV, V4332~Sgr \& \vete) was a spectral red-ward evolution connected with an emerging infrared component following the optical decay.  Here we report on the discovery of a similar feature in Keck/NIRC and {\it Spitzer} photometry  of \mef\ six\,months post-eruption. We find that its 2.1$-$22\,$\mu$m spectral energy distribution is best described by a black body with effective temperature $T_{\rm eff}=950\pm150$\,K and bolometric luminosity  $L=2.9_{-0.5}^{+0.4}\times10^{5}$\,\lsun. Assuming spherical geometry, the black body effective radius, $R=2.0_{-0.4}^{+0.6}\times10^4$\,\rsun, and corresponding expansion velocity, $v=870_{-180}^{+260}$\,km s$^{-1}$, are remarkably similar to the properties of M31~RV 70\,days after its eruption. Furthermore, we propose a search strategy for  LRNe in the local Universe making use of the longevity of their infrared excess emission and discuss the expected number of events in the {\it Spitzer} Infrared Nearby Galaxies Survey.

\end{abstract}


\section{Introduction}

\mef\ was discovered by the Lick Observatory Supernova Search on 2006 January 7. It appeared as a new transient  projected in the outskirts of  the lenticular Virgo cluster galaxy M85. The outburst, as observed in the optical, exhibited a plateau with a duration of  about  two months and reached an absolute  peak magnitude of $M_{\rm V} \cong -13$ \citep{Kulkarni:2006nt}.  Spectroscopic observations indicated that this was neither a classical nova nor a sub-luminous supernova. Instead, its early-time spectrum was consistent with a black body of $T_{\rm eff}\sim4600$\, with narrow H$\alpha$ and H$\beta$ emission. Following the plateau-phase, the optical brightness decreased rapidly and the peak emission evolved towards longer wavelengths.
 
 A similar evolution has been observed in  three other transients so far; M31RV in Andromeda  \citep{Rich:1989qy} and probably V4332~Sgr \citep{Martini:1999fk} and  the extensively studied \vete\ \citep{Brown:2002qy,Munari:2002uq,Kimeswenger:2002fj,Bond:2003kx,Evans:2003yq} in the Milky Way. 
 Among this small group, \mef\ stands out as   the most distant and  brightest event approaching a bolometric peak luminosity of  about  $5\times10^{6}$\lsun.  
 
 While the peak luminosities of the four sources differ,  their overall observed properties are indicative of  a common eruption mechanism. Thus, they were suggested to be the first members of an emerging class of eruptive transients which was recently dubbed {\it Luminous Red Novae}\footnote{The terms {\it Luminous Red Variables} \citep{Bryan:1992ly}, stars erupting into cool supergiants (SECS) \citep{Munari:2002lr} and {\it Mergebursts} \citep{Soker:2006lr} were suggested as well.} \citep[LRNe;][]{Kulkarni:2006nt}. The cause of the peculiar outbursts still needs to be fully established, although various explanations were proposed.  The most favorable scenarios appear to be  a stellar merger event in a binary system with small secondary-to-primary mass ratio \citep{Soker:2003rt,Soker:2006lr} or a planetary capture \citep{Retter:2003lr}.
 
One of the most remarkable features of M31~RV and \vete\ was a strong infrared excess which developed a  few months  after the eruption \citep{Mould:1990mz,Lynch:2004fr}\footnote{No infrared observations were obtained for V4332~Sgr within the first four years. It is thus unconstrained whether it showed a similar evolution or not.}. This excess emission is commonly associated with condensation of newly formed dust in the expelled envelope around the progenitor star. In this paper we present the results of a near and mid-infrared search for a similar component in  \mef.

\section{Observations}

We observed \mef\ using  the {\it Spitzer Space Telescope}  Infrared Array Camera \citep[IRAC;][]{Fazio:2004gf}  at 3.6, 4.5, 5.8 and 8\,$\mu$m and the peak-up imaging mode of the Infrared Spectrograph \citep[IRS;][]{Houck:2004ve} at 15.8 and 22\,$\mu$m in July 2006. See  Table~\ref{tab:log} for a summary of observations. Standard pipeline post-Basic-Calibrated-Data products were obtained from the {\it Spitzer} archive. 

\begin{figure}[htbp]
\begin{center}
\includegraphics[width=0.95\textwidth,angle=0]{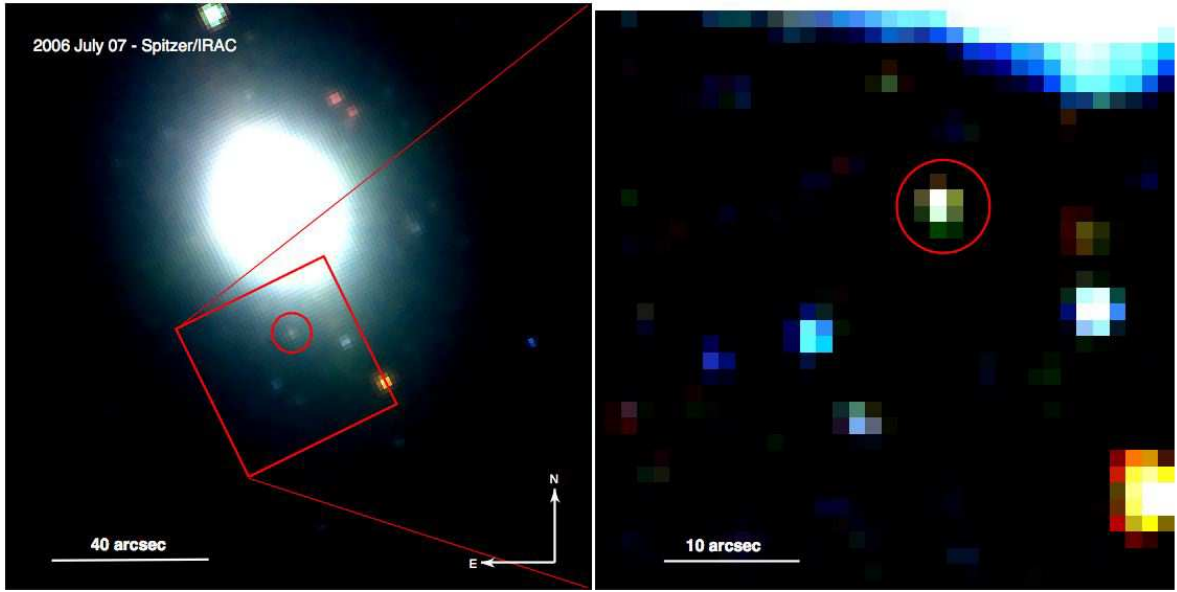}
\caption{ {\bf left:} IRAC color-composite image of M85 using 3.6, 5.8 \& 8\,$\mu$m photometry obtained $\sim6$\,months after the eruption.  The position of  \mef\  is indicated by the circle. {\bf right:}  Zoom into the location of the transient after subtraction of the galaxy light. }
\label{fig:rgb}
\end{center}
\end{figure}

A color-composite of the 3.6, 5.8 and 8\,$\mu$m emission is shown in Figure~\ref{fig:rgb} (left). Prior to undertaking photometry we subtracted the local host galaxy emission  estimated using the IRAF task {\it mkskycor} (Figure~\ref{fig:rgb} right). The flux densities are measured  through $1\farcs9$ and $3\farcs6$  apertures  for IRAC and IRS, respectively. Aperture corrections were applied according to documented values \footnote{IRAC: http://swire.ipac.caltech.edu/swire/astronomers/publications/SWIRE2\_doc\_083105.pdf, IRS: http://ssc.spitzer.caltech.edu/irs/pu\_fluxcal.txt}. See Table~\ref{tab:log} for a summary of flux densities and apparent magnitudes.

In addition, we obtained simultaneous $K$-band imaging with the Keck Observatory Near Infrared Camera \citep[NIRC;][]{Matthews:1994ul} under excellent (FWHM=0\farcs4) conditions. The data were
reduced using the near-infrared processing package {\it IRtools} for IRAF (D. Thompson, private communication). The host galaxy  emission was subtracted as described above and the flux calibration was derived using observations of the standard star SJ~9145 \cite{Persson:1998dq}. The flux measurement was obtained using a circular aperture of $0\farcs8$ radius.

\section{Results}

\subsection{$K$-band light curve}

\begin{figure}
\includegraphics[angle=-90,scale=0.6]{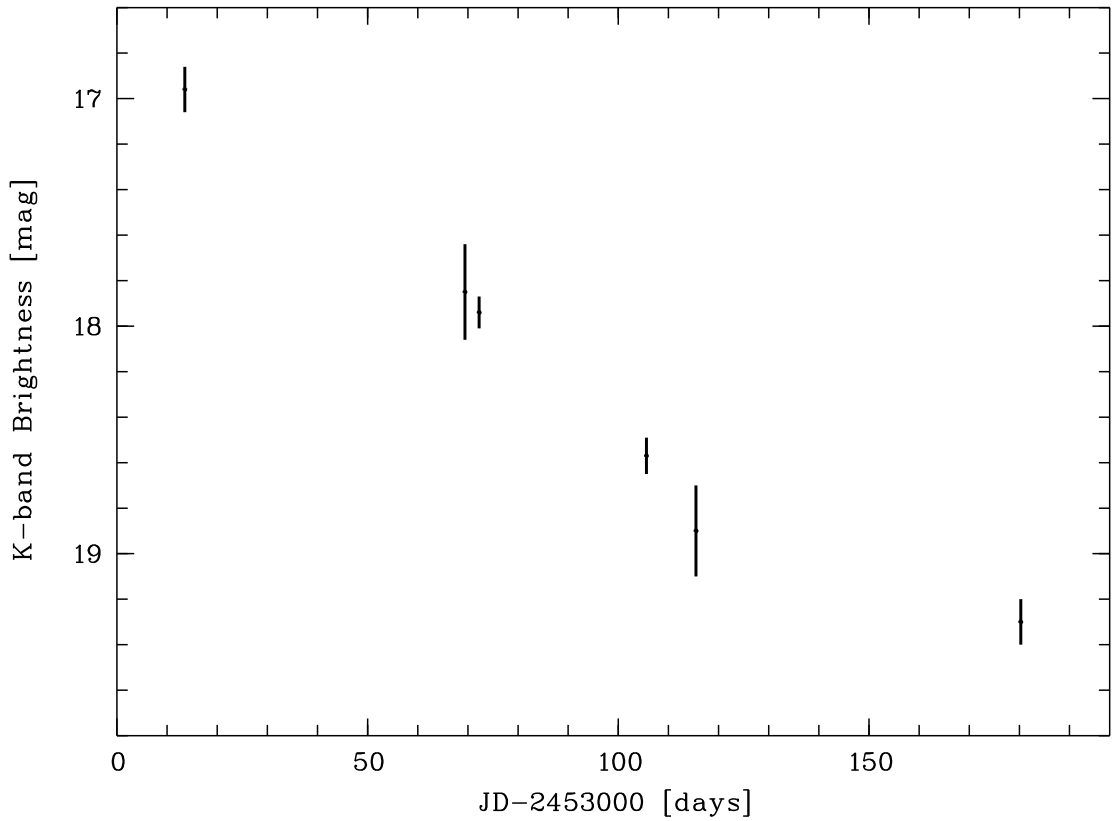}
\caption{$K$-band light curve. $t=0$ corresponds to the discovery date, 2006 January 7. Data at $t<120$\,days were reported in \citep{Kulkarni:2006nt}. The early-time light curve can be approximated by a linear decay with a slope of $\sim$0.02\,mag  per day. The late-time NIRC photometry suggests a significant  flattening between 120 and 180\,days.}
\label{fig:lc}
\end{figure}

In Figure~\ref{fig:lc} we show the $K$-band light curve combining the photometry from \cite{Kulkarni:2006nt} with the NIRC observation presented here. The sparse data are consistent with a linear decay at a rate of $\sim0.02$\,mag per day followed by a flattening at about 120$-$180\,days.

\subsection{Spectra Energy Distribution}

\begin{figure}
\includegraphics[angle=-90,scale=0.6]{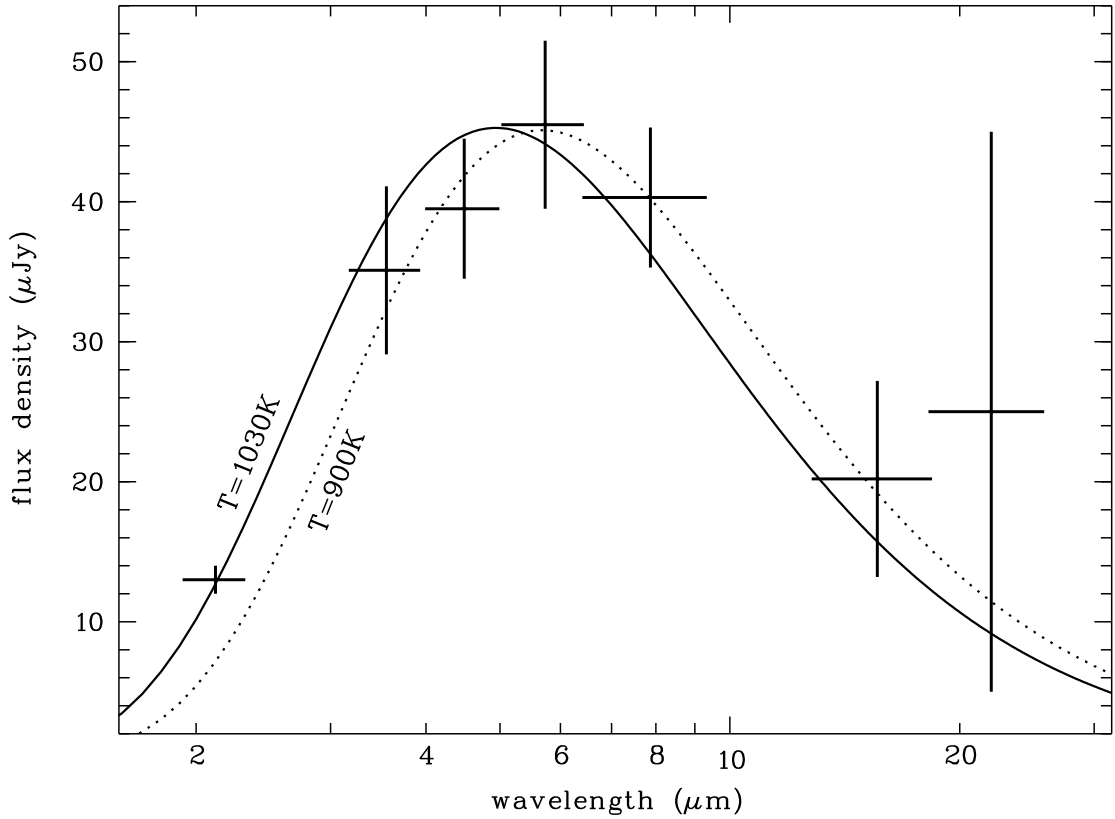}
\caption{Near and mid-infrared flux densities observed in July  2006 (crosses). The solid line shows a black body with $T_{\rm eff}=1030$\,K fitted to to 2.1$-$22\,$\mu$m photometry. Omitting the $K$-band suggests a lower $T_{\rm eff}$ (900\,K; dashed line) and an additional hotter component.}
\label{fig:sed}
\end{figure}

The 2.1 to 22$\mu$m photometry at $\sim$180\,days  is consistent with  a black body with effective temperature, $T_{\rm eff}=1030\pm50$\,K ($\chi^2/dof = 3.3/5$; solid line in Figure~\ref{fig:sed}). A fit to the IRAC and IRS data alone suggests a slightly lower temperature, $T_{\rm eff}=900_{-100}^{+140}$\,K ($\chi^2/dof = 0.9/4$; dashed  line in Figure~\ref{fig:sed}) and an additional, hotter component leading to a flux surplus at 2.1\,$\mu$m. For the following discussion we  will assume a black body temperature of $T_{\rm eff}=950\pm150$\,K for the infrared excess  emission.

The bolometric luminosity after 180\,days (as traced by $\sigma T_{\rm eff}^4$) was  $L=2.9_{-0.5}^{+0.4}\times10^{5}$\,\lsun\ (see Table~\ref{tab:bb}). The corresponding black body radius was $R=[L/(4\pi \sigma_BT_{\rm eff}^4)]^{1/2}=2.0_{-0.4}^{+0.6}\times10^4$\,\rsun. Using 2006 January  7 as the onset of the eruption,  we estimate a mean expansion velocity of  $870_{-180}^{+260}$\,km s$^{-1}$. The total emitted radiation energy is $\sim10^{47}$\,erg, dominated by the optical plateau during the first three months.

\section{Discussion and Conclusion}

We have presented the discovery of a strong 3.6--22\,$\mu$m excess in \mef\ at $\sim$180\,days.
This thermal infrared component suggests  dust condensation in the matter expelled during the eruption, similar to M31~RV \citep{Mould:1990mz} and  \vete\ \\\citep{Kimeswenger:2002fj,Lynch:2004fr}.

The derived photospheric  properties of \mef\ are resembled closest by those of M31~RV. For the latter a 1000\,K dust shell with a radius of $\sim8000$\,\rsun\ was reported after 70\,days \citep[][see Table~\ref{tab:bb} for a comparison]{Mould:1990mz}. Similar to \mef, M31~RV showed a nearly perfect black body component with  only a faint excess at shorter wavelengths. Also \vete\ displayed a single-component black body emission during the first 120\,days \citep[cf., ][]{Tylenda:2005qf}. At this time it's effective temperature, $T_{\rm eff}=3000$\,K, and radius, $R_{\rm eff}=2500$\,\rsun, differed significantly from those of \mef\ and M31~RV, though.  However, \vete\ was initially hotter ($T_{\rm eff}\sim7000$\,K) and exhibited at least two  eruption phases \citep{Munari:2002uq}. Furthermore it resides possibly in a binary  with a bright B3 V star  \citep{Munari:2002pd}, making the interpretation of the observed quantities difficult.  


Assuming  a spherical  geometry for \mef, we inferred a mean expansion velocity of  $\sim900$\,km s$^{-1}$ in June 2006. This value is surprisingly similar to the one derived for M31~RV at comparable $T_{\rm eff}$ ($\sim920$\,km s$^{-1}$, see Table~\ref{tab:bb}). However, $900$\,km s$^{-1}$ in \mef\ is  significantly larger than the velocity obtained from the  H$\alpha$ line width in February 2006 \citep[$350\pm140$\,km s$^{-1}$;][]{Kulkarni:2006nt}. A solution to this apparent discrepancy lies in the source geometry. In case of an a-spherical expansion, e.g. an oblate geometry observed face-on, the velocity ($\propto$ black body radius), may be overestimated.  Alternatively, the large velocity inferred from the infrared observations, can be caused by radiative acceleration as a result of a long lasting engine activity or multiple phases of energy injection into the expanding material. Furthermore, the Balmer emission may have originated in a low velocity region unrelated to the matter which produced the late time black body component.

In \vete\  the shell became partly transparent after $\sim$120\,days. Furthermore,  the temperature slowly increased again after $\sim$250\,days, possibly as the result of a gravitationally induced collapse of the inflated envelope \citep{Tylenda:2005qf}. A similar behavior was also reported for V4332~Sgr  about nine years after its eruption \citep{Tylenda:2005ys}. Whether or not \mef\ will evolve alike, needs to be addressed with future infrared observations. 

Most of our knowledge on LRNe prior to our analysis was based on detailed studies of a single event, \vete. Here we have shown that variations in the infrared evolution exists among the small sample of sources. Whether the cause of these variations lies in the progenitor, environment or some other unnamed parameter remains to be solved. Indeed, already the underlying stellar populations are not uniform. While \vete\ resides within a B-star cluster \citep{Afsar:2006rr}, \mef\ \citep{Kulkarni:2006nt}, M31~RV \citep{Rich:1989qy} and V4332~Sgr \citep{Tylenda:2005ys} probably originated from low-mass stars. However, it is clearly premature to speculate on a possible bimodality here.

The next step forward requires a significant increase of the LRNe sample. Estimates  based on the few known events suggest rates of approx. a dozen per year out to a distance of 20\,Mpc \citep{Kulkarni:2006nt}. Similar predictions come from the theory of violent mergers of both massive and low mass stars \citep{Soker:2006lr}. Thus, a dedicated search for LRNe is expected to net a substantial number of new events and may  revolutionize our understanding of these elusive transients.

\begin{figure}[]
\begin{center}
\includegraphics[width=0.70\textwidth,angle=-90]{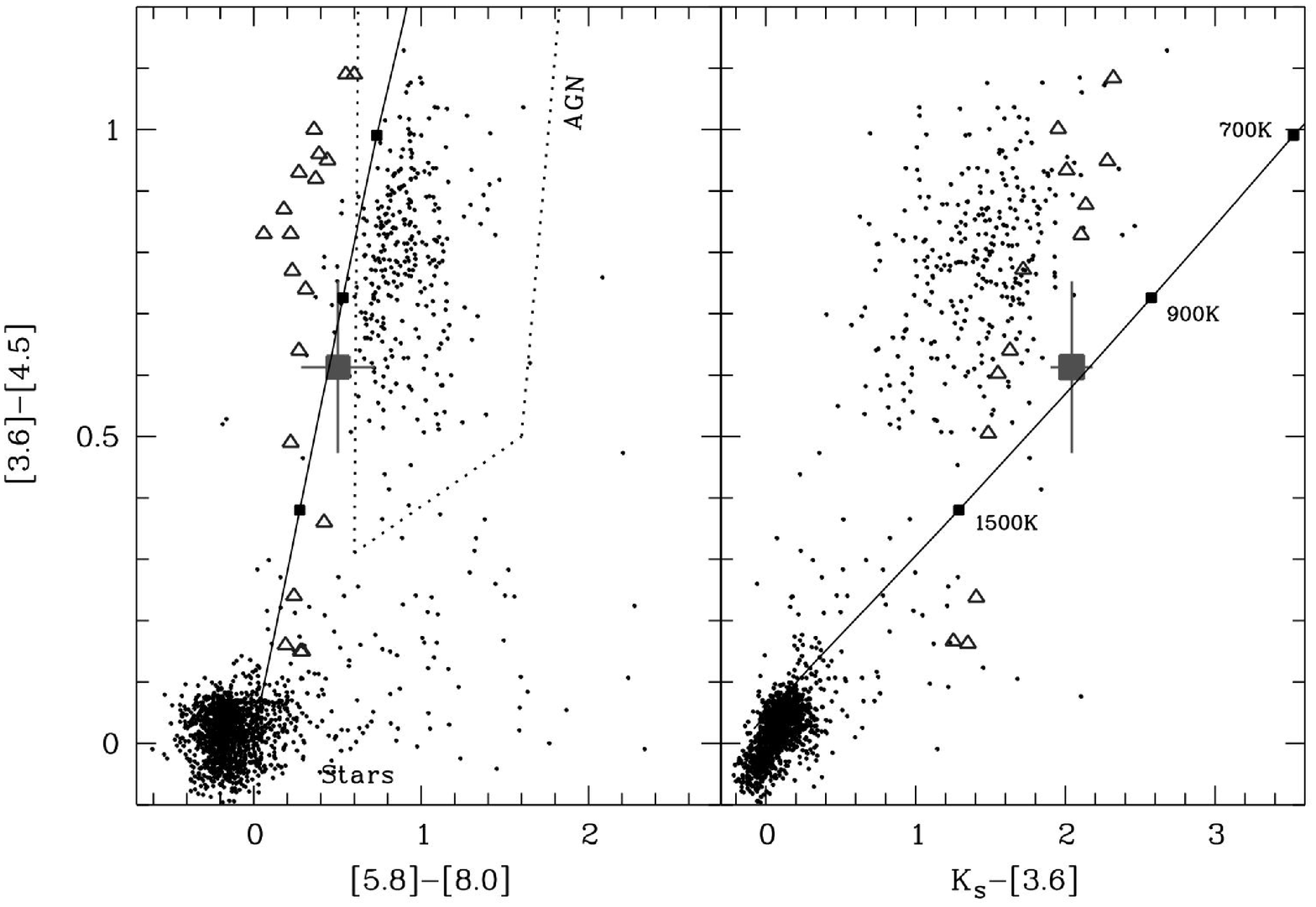}
\caption{{\bf left:} Color-color diagram of [3.6]$-$[4.5] vs. [5.8]$-$[8] 
for \mef\ (square), point-like sources in the S-COSMOS  field \citep[dots;][courtesy of the S-COSMOS team]{Sanders:2007xx} and T-dwarfs \citep[triangles;][]{Patten:2006bh}. The locations of stars and AGN \citep[contours from][]{Stern:2005lq} are indicated. Unresolved galaxies constitute the distribution of sources with [5.8]$-$[8]$>$1\,mag and [3.6]$-$[4.5]$<$0.5\,mag. Only T-dwarfs have similar colors as \mef\ after $\sim180$\,days.  Black body colors  follow the solid line. {\bf right:} Same for  [3.6]$-$[4.5] vs. $K_{\rm s}$$-$[3.6]. Strong near-infrared absorption bands in  T-dwarfs lead to a larger discrepancy compared to a pure black body spectral shape.}
\label{fig:cc}
\end{center}
\end{figure}

A promising strategy to detect new LRNe similar to \mef\ and M31~RV is to make use of their characteristic infrared emission properties. The IRAC colors of a $900-1000$\,K black body can easily be discriminated from the majority of known stellar objects, AGN and unresolved galaxies (Figure~\ref{fig:cc} left).  Only early-type T-dwarfs \citep[T2--T4;][]{Patten:2006bh}, which have similar $T_{\rm eff}$, exhibit comparable 3.6$-$8\,$\mu$m colors. However, the latter can be identified by their strong molecular absorption bands in the near-infrared \citep[e.g., CH$_4$, H$_2$0; for a recent review see][]{Kirkpatrick:2005cr}, which also lead to significant deviations from a perfect black body spectrum (offset from black body line in Figure~\ref{fig:cc}).\footnote{We note that \vete\ revealed strong absorption bands \citep[e.g., H$_2$O, CO, OH, SH;][]{Lynch:2004fr} after $\sim1$\,year together with a hot continuum ($T_{\rm eff}\sim2700$\,K) inconsistent with a simple black body. Thus, it would have been difficult to identify by its infrared colors alone. Whether this is caused by intrinsic or environmental differences with respect to \mef\ and M31~RV or only the result of a slower evolution is unconstrained.} 

The main advantage of searching in the infrared is that the detection probability is increased with respect to optical searches by  the ratio of infrared-bright time (a few years) to optical plateau duration (a few months). Using the predictions of \cite{Soker:2006lr} of approx. one LRN per galaxy every 10-50 years, we can estimate the number of anticipated events in an existing sample like the {\it Spitzer} Infrared Nearby Galaxies Survey \citep[SINGS;][]{Kennicutt:2003nx}.  SINGS comprises IRAC observations of 75 galaxies within 30\,Mpc  (mean distance 9.5\,Mpc) with sensitivities similar to the observations presented here. We predict that a search in the archival data using the described color criterion might lead to the detection of as much as 1-10 new LRN events.  A similar number of events could be found in a  large survey of nearby galaxies using the deep imaging mode of  {\it Akari} \citep[{\it ASTRO-F};][]{Matsuhara:2005eu}.

We close the paper by  emphasizing that the search for stellar mergers can be expected to have a substantial impact over the next years. Similar  to the history of the field of gamma-ray bursts,  a large increase of well studied events is anticipated to lead to exciting new insights into this emerging family of enigmatic transients. 

\acknowledgments
We are grateful to  T. Soifer, Director of the {\it Spitzer Space Telescope}, for  executing the IRAC and IRS observations during Director's Discretionary Time. {\it Spitzer} is operated by the Jet Propulsion Laboratory, California Institute of Technology, under NASA contract 1407. We thank G. Helou and L. Storrie-Lombardi for performing  the Keck/NIRC observations. We also thank M. Salvato  for valuable discussions, as well as D. Thompson for providing his near-IR data reduction routines.  This work is supported in part by grants from the National Science Foundation and NASA.

\bibliographystyle{apj}


\begin{thebibliography}{}

\bibitem[{Af{\c s}ar} \& {Bond} 2006]{Afsar:2006rr}
{Af{\c s}ar}, M. \& {Bond}, H.~E. 2006, ArXiv Astrophysics e-prints

\bibitem[{Bond}, {Henden}, {Levay}, {Panagia}, {Sparks},  {Starrfield}, {Wagner}, {Corradi}, \& {Munari} 2003]{Bond:2003kx}
{Bond}, H.~E., {Henden}, A., {Levay}, Z.~G., {et al.} 2003,  \nat, 422, 405

\bibitem[{Brown}, {Waagen}, {Scovil}, {Nelson},  {Oksanen}, {Solonen}, \& {Price} 2002]{Brown:2002qy}
{Brown}, N.~J., {Waagen}, E.~O., {Scovil}, C., {et al.} 2002, \iaucirc, 7785, 1

\bibitem[{Bryan} \& {Royer} 1992]{Bryan:1992ly}
{Bryan}, J. \& {Royer}, R.~E. 1992, \pasp, 104, 179

\bibitem[{Evans}, {Geballe}, {Rushton}, {Smalley}, {van  Loon}, {Eyres}, \& {Tyne} 2003]{Evans:2003yq}
{Evans}, A., {Geballe}, T.~R., {Rushton}, M.~T., {et al.} 2003, \mnras, 343, 1054

\bibitem[{Fazio}, {Hora}, {Allen}, {Ashby}, {Barmby},  {Deutsch}, {Huang}, {Kleiner}, {Marengo}, {Megeath}, {Melnick}, {Pahre},  {Patten}, {Polizotti}, {Smith}, {Taylor}, {Wang}, {Willner}, {Hoffmann},  {Pipher}, {Forrest}, {McMurty}, {McCreight}, {McKelvey}, {McMurray}, {Koch},  {Moseley}, {Arendt}, {Mentzell}, {Marx}, {Losch}, {Mayman}, {Eichhorn},  {Krebs}, {Jhabvala}, {Gezari}, {Fixsen}, {Flores}, {Shakoorzadeh}, {Jungo},  {Hakun}, {Workman}, {Karpati}, {Kichak}, {Whitley}, {Mann}, {Tollestrup},  {Eisenhardt}, {Stern}, {Gorjian}, {Bhattacharya}, {Carey}, {Nelson},  {Glaccum}, {Lacy}, {Lowrance}, {Laine}, {Reach}, {Stauffer}, {Surace},  {Wilson}, {Wright}, {Hoffman}, {Domingo}, \& {Cohen} 2004]{Fazio:2004gf}
{Fazio}, G.~G., {Hora}, J.~L., {Allen}, L.~E., {et al.} 2004, \apjs, 154, 10

\bibitem[{Houck}, {Roellig}, {van Cleve}, {Forrest},  {Herter}, {Lawrence}, {Matthews}, {Reitsema}, {Soifer}, {Watson}, {Weedman},  {Huisjen}, {Troeltzsch}, {Barry}, {Bernard-Salas}, {Blacken}, {Brandl},  {Charmandaris}, {Devost}, {Gull}, {Hall}, {Henderson}, {Higdon}, {Pirger},  {Schoenwald}, {Sloan}, {Uchida}, {Appleton}, {Armus}, {Burgdorf},  {Fajardo-Acosta}, {Grillmair}, {Ingalls}, {Morris}, \&  {Teplitz} 2004]{Houck:2004ve}
{Houck}, J.~R., {Roellig}, T.~L., {van Cleve}, J., {et al.} 2004, \apjs, 154, 18

\bibitem[{Kennicutt}, {Armus}, {Bendo}, {Calzetti},  {Dale}, {Draine}, {Engelbracht}, {Gordon}, {Grauer}, {Helou}, {Hollenbach},  {Jarrett}, {Kewley}, {Leitherer}, {Li}, {Malhotra}, {Regan}, {Rieke},  {Rieke}, {Roussel}, {Smith}, {Thornley}, \& {Walter} 2003]{Kennicutt:2003nx}
{Kennicutt}, Jr., R.~C., {Armus}, L., {Bendo}, {et al.} 2003, \pasp, 115, 928

\bibitem[{Kimeswenger}, {Lederle}, {Schmeja}, \&  {Armsdorfer} 2002]{Kimeswenger:2002fj}
{Kimeswenger}, S., {Lederle}, C., {Schmeja}, S., {et al.} 2002,  \mnras, 336, L43

\bibitem[{Kirkpatrick} 2005]{Kirkpatrick:2005cr}
{Kirkpatrick}, J.~D. 2005, \araa, 43, 195

\bibitem[{Kulkarni}, {Ofek}, {Rau}, {Cenko},  {Soderberg}, {Fox}, {Gal-Yam}, {Capak}, {Moon}, {Li}, {Filippenko}, {Egami},  {Kartaltepe}, \& {Sanders} 2006]{Kulkarni:2006nt}
{Kulkarni}, S.~R., {Ofek}, E.~O., {Rau}, A., {et al.} 2006,  Nature, submitted

\bibitem[{Lynch}, {Rudy}, {Russell}, {Mazuk},  {Venturini}, {Dimpfl}, {Bernstein}, {Sitko}, {Fajardo-Acosta}, {Tokunaga},  {Knacke}, {Puetter}, \& {Perry} 2004]{Lynch:2004fr}
{Lynch}, D.~K., {Rudy}, R.~J., {Russell}, R.~W., {et al.} 2004,  \apj, 607, 460

\bibitem[{Martini}, {Wagner}, {Tomaney}, {Rich},  {della Valle}, \& {Hauschildt} 1999]{Martini:1999fk}
{Martini}, P., {Wagner}, R.~M., {Tomaney}, A., {et al.} 1999, \aj, 118, 1034

\bibitem[{Matsuhara}, {Shibai}, {Onaka}, \&  {Usui} 2005]{Matsuhara:2005eu}
{Matsuhara}, H., {Shibai}, H., {Onaka}, T., {et al.} 2005, Advances in  Space Research, 36, 1091

\bibitem[{Matthews} \& {Soifer} 1994]{Matthews:1994ul}
{Matthews}, K. \& {Soifer}, B.~T. 1994, in ASSL Vol. 190: Astronomy with  Arrays, The Next Generation, ed. I.~S. {McLean}, 239--+

\bibitem[{Mould}, {Cohen}, {Graham}, {Hamilton},  {Matthews}, {Picard}, {Reid}, {Schmidt}, {Soifer}, {Wilson}, {Rich}, \&  {Gunn} 1990]{Mould:1990mz}
{Mould}, J., {Cohen}, J., {Graham}, J.~R., {et al.} 1990, \apjl, 353, L35

\bibitem[{Munari}, {Desidera}, \&  {Henden} 2002a]{Munari:2002pd}
{Munari}, U., {Desidera}, S., \& {Henden}, A. 2002a, \iaucirc,  8005, 2

\bibitem[{Munari}, {Henden}, {Kiyota},  {Laney}, {Marang}, {Zwitter}, {Corradi}, {Desidera}, {Marrese}, {Giro},  {Boschi}, \& {Schwartz} 2002b]{Munari:2002uq}
{Munari}, U., {Henden}, A., {Kiyota}, S., {et al.} 2002b, \aap, 389, L51

\bibitem[{Patten}, {Stauffer}, {Burrows}, {Marengo},  {Hora}, {Luhman}, {Sonnett}, {Henry}, {Raghavan}, {Megeath}, {Liebert}, \&  {Fazio} 2006]{Patten:2006bh}
{Patten}, B.~M., {Stauffer}, J.~R., {Burrows}, A., {et al.} 2006, \apj, 651, 502

\bibitem[{Persson}, {Murphy}, {Krzeminski}, {Roth}, \&  {Rieke} 1998]{Persson:1998dq}
{Persson}, S.~E., {Murphy}, D.~C., {Krzeminski}, W., {et al.} 1998, \aj, 116, 2475

\bibitem[{Rich}, {Mould}, {Picard}, {Frogel}, \&  {Davies} 1989]{Rich:1989qy}
{Rich}, R.~M., {Mould}, J., {Picard}, A., {et al.} 1989, \apjl, 341, L51

\bibitem[{Sanders}, {Salvato}, {Aussel}, {Ilbert},  {Scoville}, {Surace}, {Frayer}, {Sheth}, {Helou}, {Brooke}, {Bhattacharya},  {Yan}, {Kartaltepe}, {Barnes}, {Blain}, {Calzetti}, {Capak}, {Carilli},  {Carollo}, {Comastri}, {Daddi}, {Ellis}, {Elvis}, {Fall}, {Franceschini},  {Giavalisco}, {Hasinger}, {Impey}, {Koekemoer}, {Le F\`evre}, {Lilly}, {Liu},  {McCracken}, {Mobasher}, {Renzini}, {Rich}, {Schinnerer}, {Shopbell},  {Taniguchi}, {Thompson}, {Urry}, \& {Williams} 2007]{Sanders:2007xx}
{Sanders}, D.~B., {Salvato}, M., {Aussel}, H., {et al.} 2007, ApJS, submitted

\bibitem[{Soker} \& {Tylenda} 2003]{Soker:2003rt}
{Soker}, N. \& {Tylenda}, R. 2003, \apjl, 582, L105

\bibitem[{Soker} \& {Tylenda} 2006]{Soker:2006lr}
---. 2006, \mnras, 1202

\bibitem[{Stern}, {Eisenhardt}, {Gorjian}, {Kochanek},  {Caldwell}, {Eisenstein}, {Brodwin}, {Brown}, {Cool}, {Dey}, {Green},  {Jannuzi}, {Murray}, {Pahre}, \& {Willner} 2005]{Stern:2005lq}
{Stern}, D., {Eisenhardt}, P., {Gorjian}, V., {et al.} 2005, \apj, 631, 163

\bibitem[{Tylenda} 2005]{Tylenda:2005qf}
{Tylenda}, R. 2005, \aap, 436, 1009

\bibitem[{Tylenda}, {Crause}, {G{\'o}rny}, \&  {Schmidt} 2005]{Tylenda:2005ys}
{Tylenda}, R., {Crause}, L.~A., {G{\'o}rny}, S.~K., {et al.} 2005,  \aap, 439, 651

\end{thebibliography}

\clearpage
\begin{table}
\begin{center}
\caption{Log of observations.\label{tab:log}}
\begin{tabular}{cccccc}
\tableline\tableline
Start UTC & Instrument & Filter & Exposure  & Flux density & Brightness\\
		&		     & [$\mu$m] & [s] & [$\mu$Jy] & [Vega mag]\\
\tableline
2006 July 6.28 & Keck/NIRC & 2.1 & 1200 & 13$\pm$1& 19.3$\pm$0.1\\
2006 July 7.06 & Spitzer/IRAC & 3.6 & 3000 & 35$\pm$6 & 17.2$\pm$0.1 \\
2006 July 7.06 & Spitzer/IRAC & 4.5 & 3000 & 40$\pm$5 & 16.6$\pm$0.1\\
2006 July 7.06 & Spitzer/IRAC & 5.8 & 3000 & 46$\pm$6 & 16.0$\pm$0.1\\
2006 July 7.06 & Spitzer/IRAC & 8.0 & 3000 & 40$\pm$5 & 15.5$\pm0.1$\\
2006 July 2.12 & Spitzer/IRS & 15.6 & 3000 & 20$\pm$7 & 14.9$^{+0.5}_{-0.3}$\\
2006 July 2.12 & Spitzer/IRS & 22.0 & 3000 & 25$\pm$20 & 14.0$^{+1.8}_{-0.7}$\\
\tableline
\end{tabular}
\end{center}
\end{table}

\begin{table}
\begin{center}
\caption{Inferred black body parameters.\label{tab:bb}}
\begin{tabular}{lccccccc}
\tableline\tableline
Source &$L_{\rm peak}$ & $T_{\rm eff, peak}$ & $R_{\rm peak}$ & $L_{\rm late\tablenotemark{a}}$ & $T_{\rm eff, late}$ 	& $R_{\rm late}$  & $R_{\rm late}/t$\\
&[$\times10^5$\,\lsun]&  [$\times10^3$\,K]	&[$\times10^3$\,\rsun] & [$\times10^5$\,\lsun] & [$\times10^3$\,K] & [$\times10^3$\,\rsun] & [km s$^{-1}$]\\

\tableline
\mef\tablenotemark{b} &	$\sim50$ & $\sim4.6$	&	$\sim3.6$	& $2.9_{-0.5}^{+0.4}$& $0.95\pm0.15$	&	$20_{-4}^{+6}$ & 870$_{-180}^{+260}$\\
M31~RV\tablenotemark{c} & $\sim8$ & $\sim4$	& $\sim2$ 	&	$\sim0.6$ &$\sim1$ & $\sim8$ & $\sim920$\\

\tableline
\end{tabular}
\tablenotetext{a}{ at $t\sim180$\,days for \mef\ and $t\sim70$\,days for M31~RV}
\tablenotetext{b}{ peak values from \cite{Kulkarni:2006nt}. Late time values this paper}
\tablenotetext{c}{ peak luminosity from \cite{Rich:1989qy}. Remaining values from \cite{Mould:1990mz}}

\end{center}
\end{table}

\clearpage

\end{document}